\documentclass[showpacs,aps,twocolumn]{revtex4}

\usepackage{bm}
\usepackage{amsmath}
\usepackage{graphicx}
\usepackage{subfigure}
\usepackage[usenames,dvipsnames]{color}
\definecolor{darkblue}{RGB}{0,0,196}
\usepackage[colorlinks=true,linkcolor=darkblue,citecolor=darkblue,urlcolor=darkblue]{hyperref}

\usepackage{setspace}
\usepackage{footmisc}
\usepackage[makeroom]{cancel}
\usepackage{comment}

\usepackage{color,soul}

\def\be{\begin{equation}}
\def\ee{\end{equation}}
\def\ba{\begin{eqnarray}}
\def\ea{\end{eqnarray}}

\usepackage{graphicx}
\usepackage{amsmath,bbm}
\usepackage{amssymb,bm}

\begin{document}

\title{Electrical conductivity of Hot and Dense QCD matter created in Heavy-Ion Collisions: A Color String Percolation Approach}
\author{Pragati Sahoo}
\author{Swatantra~Kumar~Tiwari}
\email{sktiwari4bhu@gmail.com}
\author{Raghunath~Sahoo\footnote{Corresponding author: $Raghunath.Sahoo@cern.ch$}}
\affiliation{Discipline of Physics, School of Basic Sciences, Indian Institute of Technology Indore, Indore- 453552, INDIA}

\begin{abstract}
\noindent

Recently, transport coefficients viz. shear viscosity, electrical conductivity etc. of strongly interacting matter produced in heavy-ion collisions have drawn considerable interest. We study the normalised electrical conductivity ($\sigma_{\rm el}$/T) of hot QCD matter as a function of temperature (T) using the Color String Percolation Model (CSPM). We also study the temperature dependence of shear viscosity and its ratio with electrical conductivity for the QCD matter. We compare CSPM estimations with various existing results and lattice Quantum Chromodynamics (lQCD) predictions with (2+1) dynamical flavours. We find that $\sigma_{\rm el}$/T in CSPM has a very weak dependence on the temperature. We compare CSPM results with those obtained in Boltzmann Approach to Multi-Parton Scatterings (BAMPS) model. A good agreement is found between CSPM results and predictions of BAMPS with fixed strong coupling constant.
\end{abstract}


\date{\today}

\maketitle 
\section{Introduction}
\label{intro}
Ultra-relativistic heavy-ion collision programs at Relativistic Heavy-Ion Collider (RHIC) and Large Hadron Collider (LHC) produce a strongly interacting matter known as Quark-Gluon Plasma (QGP)~\cite{Gyulassy:2004zy}. Various experimental studies have been done in order to characterise the properties and behaviour of matter at extreme conditions of temperature and energy densities. The transport properties are very important to understand the evolution of the strongly interacting matter produced in heavy-ion collisions. These are mainly the theoretical inputs to the hydrodynamical calculations and affect various observables such as elliptic flow, transverse momentum spectra of particles created in heavy-ion collisions~\cite{Gale:2013da,Schenke:2011zz,Heinz:2013th}. A very small shear viscosity to entropy density ratio explains the elliptic flow of identified hadrons produced at RHIC and LHC energies \cite{starWhite}  and suggests the fluidity of the hot QCD matter produced. Various methods are used to estimate the shear viscosity ($\eta$) such as Kubo formalism~\cite{Kubo:1957mj}, effective models~\cite{Plumari:2012ep,Sasaki:2008fg,Dobado:2008vt,Chakraborty:2010fr,Zhuang:1995uf,Wiranata:2012br,Ghosh:2014ija} etc. 

Electrical conductivity ($\sigma_{\rm el}$) is another key transport coefficient in order to understand the behaviour and properties of strongly interacting matter. This plays an important role in the hydrodynamic evolution of the matter produced in heavy-ion collisions where charge relaxation takes place. In ref.~\cite{Hirono:2012rt}, the electrical conductivity is extracted from charge dependent flow parameters from asymmetric heavy ion collisions. Experimentally, it has been observed that very strong electric and magnetic fields are created in the early stages (1-2 fm/c) of non-central collisions of nuclei at RHIC and LHC ~\cite{Hirono:2012rt,Tuchin:2013ie}. The values of the electric and magnetic fields at RHIC are eE $\approx m_\pi^2 \approx 10^{\rm 21}$ V/cm and eB $\approx m_\pi^2 \approx 10^{\rm 18}$ G~\cite{Tuchin:2013ie}. Such a large electrical field influences the medium, which depends on the electrical conductivity. $\sigma_{\rm el}$ is responsible for producing an electric current in the early stage of the heavy-ion collision. 

Although with the prior knowledge of color charges and the associated electric charges of the quarks, one might presume the QCD matter to be highly conductive. In  contrast, this assumption fails due to the high interaction rates of the produced QCD matter, which again suggests low shear viscosity to entropy density ratio ($\eta$/s). In highly conducting quark-gluon plasma, the screening of external electromagnetic fields happens due to the high values of $\sigma_{\rm el}$ like the Meissner effect in superconductors as well as the ``skin effect" for the electric current~\cite{Cassing:2013iz}. The electrical conductivity is one of the fundamental reasons for chiral magnetic effect~\cite{Fukushima:2008xe}, which is a signature of CP violation in the strong interaction. In view of this, a detailed study of electrical conductivity in the strongly interacting QCD matter is inevitable.

The experimental measurement of electrical conductivity ($\sigma_{\rm el}$) of the matter produced in heavy-ion collisions is not possible. Its information can be extracted from flow parameters measured in heavy-ion collision experiments~\cite{Hirono:2012rt}. Recently, various theoretical approaches have been used to study the electrical conductivity~\cite{Arnold:2000dr,Arnold:2003zc,Gupta:2003zh,Aarts:2007wj,Buividovich:2010tn,Ding:2010ga,Burnier:2012ts,Brandt:2012jc,Amato:2013naa,Cassing:2013iz,Steinert:2013fza,Puglisi:2014pda,Finazzo:2013efa,Mitra:2016zdw,Srivastava:2015via}. $\sigma_{\rm el}$ is also related to the soft dilepton production rate~\cite{Moore:2006qn} and the magnetic field diffusion in the medium~\cite{Baym:1997gq,FernandezFraile:2005ka}. 

Color String Percolation Model is a QCD inspired model~\cite{Armesto:1996kt,Nardi:1998qb,Braun:1999hv,Braun:1997ch,Braun:2000hd}, which can be used as an alternative approach to Color Glass Condensate (CGC). In CSPM, the color flux tubes are stretched between the colliding partons in terms of the color field. The strings produce $\it q\bar q$ pair in finite space filled similarly as in the Schwinger mechanism of pair creation in a constant electric field covering all the space~\cite{Phyreport}. With the growing energy and the number of nucleons of participating nuclei, the number of strings grows. Color strings may be viewed as small discs in the transverse space filled with the color field created by colliding partons. The number of strings grows as energy and size of the colliding nuclei increase and starts overlapping to form clusters. After a critical string density reached, a macroscopic cluster appears that marks the percolation phase transition which spans the transverse nuclear interaction area. 2D percolation is a non-thermal second order phase transition. In CSPM, the Schwinger barrier penetration mechanism for particle production, the fluctuations in the associated string tension and the quantum fluctuations of the color fields make it possible to define a temperature. Consequently, the particle spectrum is produced with a thermal distribution. When the initial density of interacting colored strings ($\xi$) exceeds the 2D percolation threshold ($\xi_c$) i.e. $\xi > \xi_c $, a macroscopic cluster appears, which defines the onset of color deconfinement. The critical density of percolation is related to the effective critical temperature and thus percolation may be a possible way to achieve deconfinement in ultrarelativistic heavy-ion collisions \cite{PLB642} and in high multiplicity pp collisions~\cite{Gutay:2015cba,Hirsch:2018pqm}. It is observed that, CSPM can be successfully used to describe the initial stages in high energy heavy-ion collisions \cite{Phyreport}. Recently, we have performed collision centrality, energies and species dependent study of the deconfinement phase transition at RHIC Beam Energy Scan (BES) energies using color string percolation model~\cite{Sahoo:2018dcz}. We have also studied various thermodynamical and transport properties at RHIC BES energies in this approach~\cite{Sahoo:2017umy}.

In this work, for the first time we give the formulation of $\sigma_{\rm el}$ in the color string percolation approach. The paper is organised as: In section~\ref{elec}, we give the detailed formulation for calculation of electrical conductivity and shear viscosity in CSPM and present results and discussions in section~\ref{RD}. Finally, we present summary and conclusions in section~\ref{summary}.

\section{Electrical Conductivity and Shear Viscosity}
\label{elec}
In this section, we develop the formulation for evaluating the electrical conductivity of strongly interacting matter using the color string percolation approach. We start with few basic equations of CSPM. The percolation density parameter, $\xi$ for central Au+Au collisions at RHIC energies is calculated by using the parameterisation of pp collisions at $\sqrt{s}$ = 200 GeV as discussed below. In CSPM one obtains:

\begin{eqnarray}
\frac{dN_{\rm ch}}{dp_{\rm T}^{2}} = \frac{a}{(p_{\rm 0}+{p_{\rm T}})^{\alpha}},
\end{eqnarray}

where, a is the normalisation factor and  $p_{0}$, $\alpha$ are fitting parameters given as, $p_{0}$ = 1.982 and $\alpha$ = 12.877 \cite{Braun:2015eoa}. Due to the low string overlap probability in pp collisions the fit parameters are then used to evaluate the interactions of the strings in Au+Au collisions as,

\begin{eqnarray}
 p_{\rm 0}\rightarrow p_{\rm 0}\left(\frac{\langle  nS_{\rm 1}/S_{\rm n}\rangle_{\rm Au+Au}}{\langle nS_{\rm 1}/S_{\rm n}\rangle_{\rm pp}}\right)^{1/4}.
 \label{po}
\end{eqnarray}

Here, $S_{\rm n}$ corresponds to the area occupied by $n$ overlapping strings. Now,  

\begin{eqnarray}
 \langle \frac{nS_{\rm 1}}{S_{\rm n}} \rangle = \frac{1}{F^{\rm 2}(\xi)},
 \label{p1}
\end{eqnarray}

where, $F(\xi)$ is the color suppression factor, which is given as,

\begin{eqnarray}
 F(\xi) = \sqrt \frac{1-e^{-\xi}}{\xi}.
 \label{p1}
\end{eqnarray}

To calculate the electrical conductivity of strongly interacting matter, which is one of the most important transport properties of QCD matter, we proceed as follows. The mean free path, which describes the relaxation of the system far from equilibrium can be written in terms of number density and cross-section as,

\begin{eqnarray}
\lambda_{\rm mfp} = \frac{1}{n\sigma_{\rm tr}},
\label{eq5}
\end{eqnarray}

where n is the number density of an ideal gas of quarks and gluons and $\sigma_{\rm tr}$ is the transport cross-section.
In CSPM the number density is given by the effective number of sources per unit volume

\begin{eqnarray}
n = \frac{N_{\rm sources}}{S_{\rm n}L}.
\label{el}
\end{eqnarray}

Here, $L$ is the longitudinal extension of the string $\sim $1 fm. The area occupied by the strings is given by the relation $(1 - e^{-\xi})S_{\rm n}$. Thus, the effective number of sources is given by the total area occupied by the strings divided by the area of an effective string, $S_1F({\xi})$ as shown below,
 
\begin{eqnarray}
N_{\rm sources} = \frac{(1 - e^{-\xi})S_{\rm n}}{S_{\rm 1}F({\xi})},
\label{el}
\end{eqnarray}

In general, $N_{\rm sources}$ is smaller than the number of single strings. $N_{\rm sources}$ equals to the number of strings $N_{\rm s}$ in the limit of $\xi = 0$. So,

\begin{eqnarray}
n = \frac{(1 - e^{-\xi})}{S_{\rm 1}F({\xi})L}.
\label{eq8}
\end{eqnarray}

Now, using eqs.~\ref{eq5} and~\ref{eq8}, we get,

\begin{eqnarray}
\lambda_{\rm mfp} = \frac{L}{(1 - e^{-\xi})},
\label{mfp}
\end{eqnarray}

where $\sigma_{\rm tr}$, the transverse area of the effective strings equals to $S_1F(\xi)$.  

\begin{figure}
\includegraphics[height=25em]{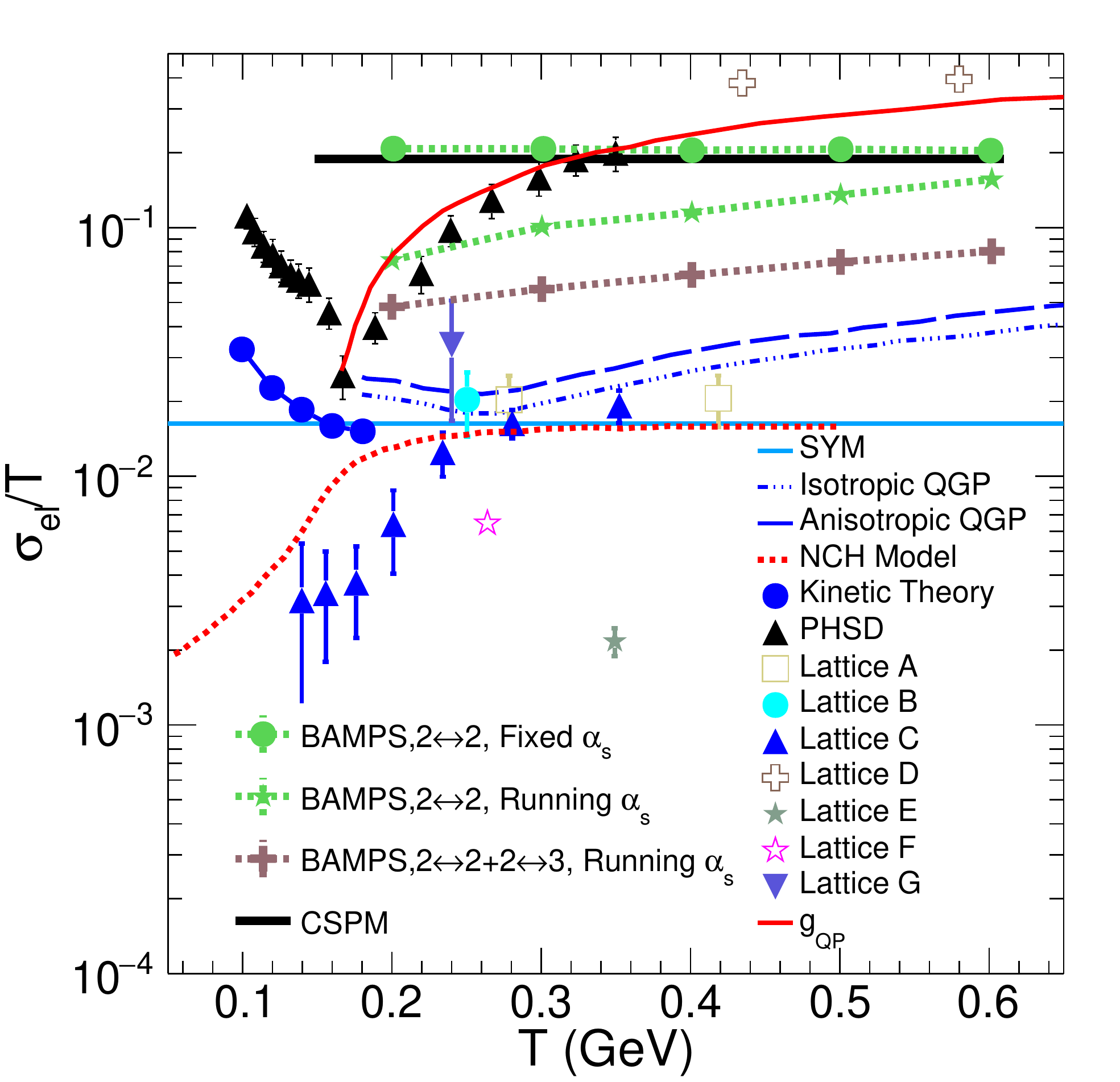}
\caption[]{(colour online) $\sigma_{\rm el}$/T versus T plot. The black solid line is the result obtained in CSPM and black triangles are PHSD results~\cite{Cassing:2013iz}. The green and brown dotted lines correspond to various BAMPS results~\cite{Greif:2014oia}. The NCH model~\cite{Finazzo:2013efa} results are shown by the red dotted line. The blue circles are kinetic theory calculations~\cite{Greif:2016skc}. The horizontal line is the result obtained for conformal supersymmetric (SYM) Yang-Mills Plasma~\cite{CaronHuot:2006te}. Lattice data: lattice A- G~\cite{Aarts:2007wj,Brandt:2012jc,Amato:2013naa,Gupta:2003zh,Buividovich:2010tn,Burnier:2012ts,Ding:2010ga,Aarts:2014nba} are also shown by various symbols in the figure. The results for isotropic and anisotropic QGP~\cite{Thakur:2017hfc} are shown by the blue dash-dotted and dashed lines, respectively. The red solid line depicts the results of quasi-particle (QP) model~\cite{Puglisi:2014pda}.}
\label{EC}
\end{figure}

Now we derive the formula for electrical conductivity. For this, we use Anderson-Witting model, in which the Boltzmann transport equation is given as~\cite{Anderson},

\begin{eqnarray}
p^\mu\partial_\mu f_k + qF^{\alpha\beta}p_{\beta}\frac{\partial f_k}{\partial p^{\alpha}} = \frac{-p^\mu u_{\mu}}{\tau}(f_k - f_{eq,k}),
\label{bte}
\end{eqnarray}

where $f_k = f(x,\overrightarrow{p},t)$ is the full distribution function and $f_{eq,k}$ is the equilibrium distribution function of $\rm k^{th}$ species. $\tau$ is the mean time between collisions and $u_\mu$ is the fluid four velocity in the local rest frame. Eq.~\ref{bte} provides a straightforward calculation of the quark distribution after applying the electric field. The gluon distribution function remains thermal and not altered by electric field. Here, we assume that there are as many quarks (charge $q$) as anti-quarks (charge -$q$) and uncharged gluons in the system. $F^{\alpha\beta}$ is the electromagnetic field strength tensor, which in terms of electric field and the magnetic flux tensor is given as~\cite{Greif:2014oia},

\begin{eqnarray}
F^{\mu\nu} = u^{\nu} E^{\mu} - u^{\mu} E^{\nu} - B^{\mu\nu}.
\label{field}
\end{eqnarray}
Since we study the effect of electric field, the magnetic field is set to zero, $B^{\mu\nu} = 0$ in the calculations. The electric current density of the $\rm k^{th}$ species in the $x$-direction is given as,

\begin{eqnarray}
j^x_k = q_k\int \frac{d^3p p^x}{(2\pi)^3 p^0}f_k = g_k\tau\frac{8}{3}\frac{\pi q_{k}^2 T^2}{(2\pi)^3}E^x.
\label{Current}
\end{eqnarray}
According to Ohm's law, $j^x_k = \sigma_{el}E^x$. Using eq.~\ref{Current} and relation $n_k = g_k T^3/\pi^2$, electrical conductivity in the assumption of very small electric field and no cross effects between heat and electrical conductivity in the relaxation time approximation is given by,

\begin{eqnarray}
\sigma_{\rm el} = \frac{1}{3T} \sum_{\rm k=1}^{M}q_{\rm k}^2 n_{\rm k} \lambda_{\rm mfp}.
\label{el_con}
\end{eqnarray}

Putting eq.~\ref{mfp} in eq.~\ref{el_con} and considering the density of up quark$(u)$ and its antiquark$(\bar{u})$ in the calculation, we get the expression for $\sigma_{\rm el}$ as, 

\begin{eqnarray}
\sigma_{\rm el} = \frac{1}{3T}\frac{4}{9}e^{2}n_{\rm q}(T) \frac{L}{(1 - e^{-\xi})}.
\label{el_c}
\end{eqnarray}

Here, the pre-factor 4/9 reflects the fractional quark charge squared $( \sum_{\rm f} q_{\rm f}^{2})$ and $n_{\rm q}$ denotes the total density of quarks or antiquarks. Here, $e^2$ in the natural unit is taken as $4\pi\alpha$, where $\alpha$ = 1/137.

In the framework of a relativistic kinetic theory, the shear viscosity over entropy density ratio, $\eta/s$ is given by \cite{DiasdeDeus:2012uc,Danielewicz:1984ww,Hirano:2005wx},

\begin{eqnarray}
 \eta/s \simeq \frac{T\lambda_{mfp}}{5},
 \label{e1}
\end{eqnarray}

In the context of CSPM the above equation can be reduced using eq.~\ref{mfp} as,

\begin{eqnarray}
 \eta/s \simeq \frac{TL}{5(1-e^{-\xi})}.
 \label{e3}
\end{eqnarray}

\begin{figure}
\includegraphics[height=25em]{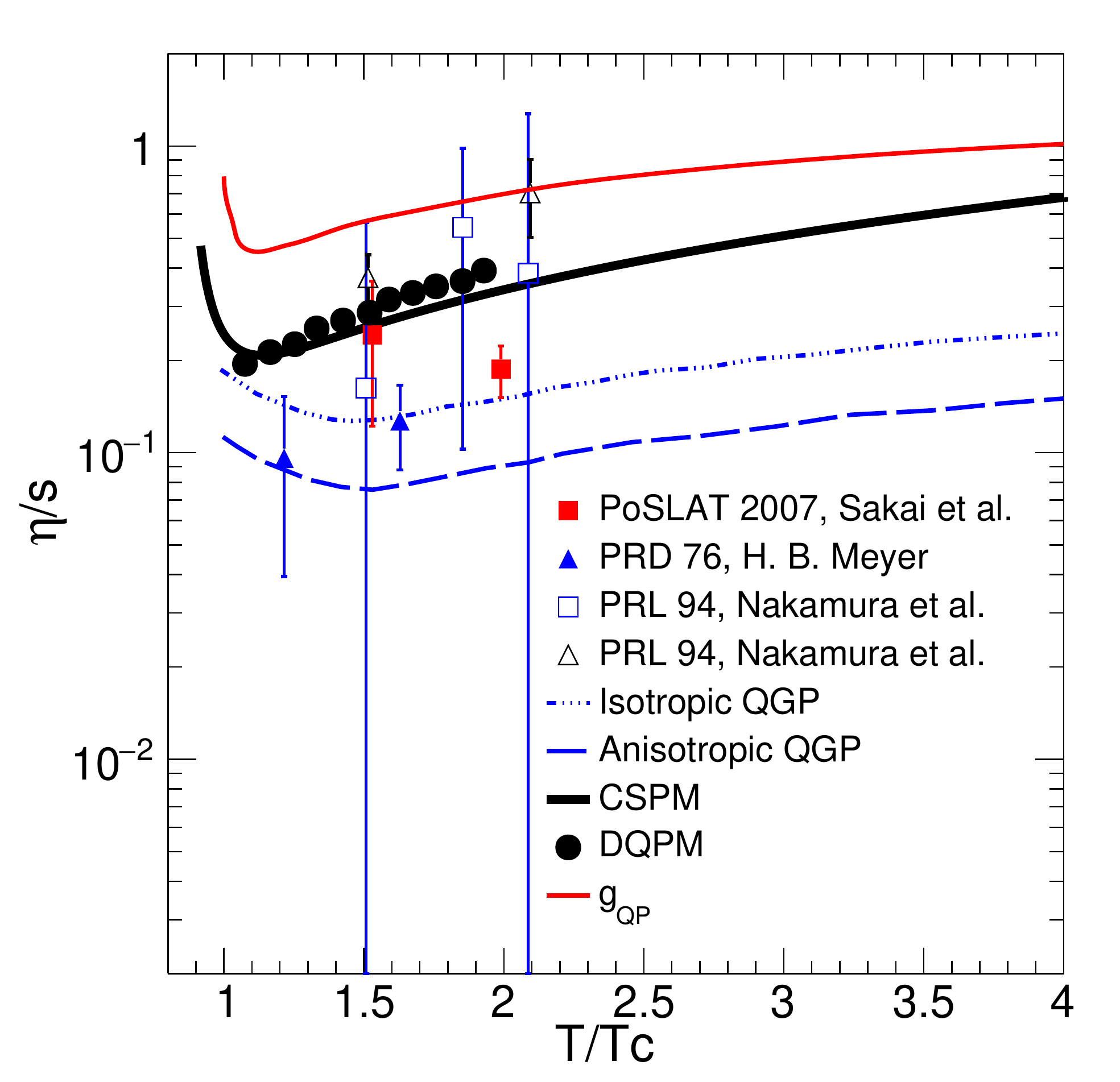}
\caption[]{(colour online) The ratio $\eta$/s as a function of $\rm T/T_c$. The black solid line is the CSPM result and broken lines are results from ref.~\cite{Thakur:2017hfc}. The symbols are lattice QCD results: full triangles~\cite{Meyer:2007ic}, open squares and open triangles~\cite{Nakamura:2004sy}, full squares~\cite{Sakai:2007cm}. The black circles are the results obtained in DQPM~\cite{Marty:2013ita}. The red solid line is QP model results~\cite{Puglisi:2014pda}.}
\label{eta_s}
\end{figure}

\begin{figure}
\includegraphics[height=25em]{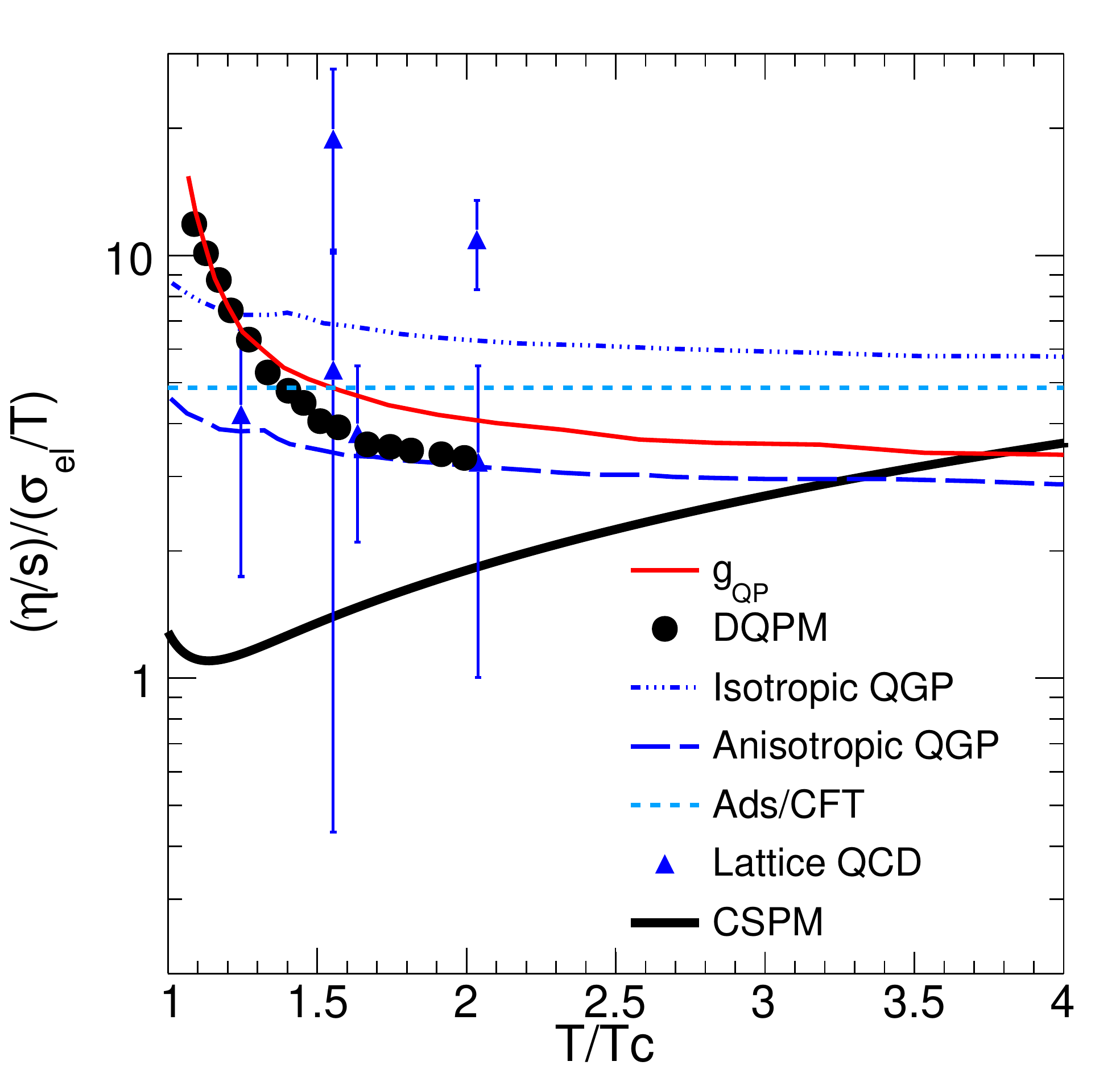}
\caption[]{(colour online) The ratio $\eta$/s and $\sigma_{\rm el}$/T with respect to $T/T_{c}$. The black solid line is the CSPM result and broken lines are results from ref.~\cite{Thakur:2017hfc}. The symbols are lattice QCD results~\cite{Puglisi:2014pda}. The DQPM and QP results are shown by the black circles and red solid line, respectively~\cite{Puglisi:2014pda}.}
\label{el}
\end{figure}

\section{Results and Discussions}
\label{RD}
In this section, we discuss the results obtained in CSPM along with those obtained in various approaches. In fig.~$\ref{EC}$, we show $\sigma_{\rm el}$/T as a function of temperature. The lQCD estimations i.e. lattice A - G~\cite{Aarts:2007wj,Brandt:2012jc,Amato:2013naa,Gupta:2003zh,Buividovich:2010tn,Burnier:2012ts,Ding:2010ga,Aarts:2014nba} are shown in the figure for comparison. The green and brown dotted lines are the result of microscopic transport model BAMPS~\cite{Greif:2014oia}, in which the relativistic (3+1)- dimensional Boltzmann equation is solved numerically to extract the electric conductivity for a dilute gas of massless and classical particles described by the relativistic Boltzmann equation. The green dotted line with the solid circles is the result for only elastic processes 2$\leftrightarrow$2, where strong coupling constant ($\alpha_s$) is taken as constant ($\alpha_s$ = 0.3) and the green dotted line with the solid stars is with the same setup for running $\alpha_s$. The brown dotted line with the brown plus symbols is the BAMPS result, where both elastic 2$\leftrightarrow$2 and inelastic 2$\leftrightarrow$3 processes are taken into consideration with running $\alpha_s$. The  BAMPS results show a slower increase of $\sigma_{\rm el}$/T with temperature for both the cases of running $\alpha_s$ as the effective cross section changes with the temperature, while $\sigma_{\rm el}$/T remains almost independent of temperature for the case of constant $\alpha_s$. The BAMPS results are above the lQCD results. The solid black line shows our results of CSPM for $u$- quark and antiquark calculated using eq.~\ref{el_c}. We observe that $\sigma_{\rm el}/T$ is almost independent of temperature and matches with the results of BAMPS with constant $\alpha_s$, which may be due to the similar basic ingredients and procedure for the estimation of $\sigma_{\rm el}/T$.  

Although the percolation of string approach is not directly obtained from QCD but it is QCD inspired, as like the BAMPS model is governed by pQCD. The basic ingredients for the percolation are strings, which are stretched between the partons of the projectile and target and forms color electric and magnetic field in the longitudinal directions. The color strings fragment into $q-\bar{q}$ and/or $qq-\bar{qq}$ pairs and form hadrons~\cite{Phyreport}. In the present study, we consider the strings to fragment into only  $u-\bar{u}$ pairs. We use Drude formula in the relativistic case to estimate the electrical conductivity, which can be obtained after solving the relativistic Boltzmann transport equation with some approximations as mentioned in the formulation section. So, the observation proclaims the almost similar approach of both the models for the calculation of $\sigma_{\rm el}/T$. It has been shown in ref.~\cite{Puglisi:2014sha} that the real electrical conductivity can be even more than a 50\% larger than the estimate of the Drude formula unless the cross section is isotropic (no angular dependence).

 A non-conformal holographic model~\cite{Finazzo:2013efa} is used to estimate the electrical conductivity of the strongly coupled QGP, which is shown by the red dashed line and explains the lQCD data qualitatively. Kinetic theory~\cite{Greif:2016skc} is also used to calculate electrical conductivity of hadron gas whose results are shown by blue circles in the figure, which shows a decrease of $\sigma_{\rm el}$/T with temperature. The electrical conductivity for conformal Yang-Mills plasma~\cite{CaronHuot:2006te} is also shown by the horizontal line in the figure. The blue dash-dotted and dotted lines are the results for QGP obtained using the quasi-particle model for quark and gluons~\cite{Thakur:2017hfc} for isotropic and anisotropic cases, respectively. Here, all the quarks and antiquarks have both the masses i.e. thermal and bare. The thermal masses of quarks and antiquarks arise due to the interaction with the constituents of the medium. Parton-Hadron-String Dynamics (PHSD) model results~\cite{Cassing:2013iz} are also shown by the black triangles in the figure for both the phases- hadron gas and quark-gluon plasma with different approaches. The hadron-string-dynamics transport approach has been used for the hadronic sector PHSD, while the partonic dynamics in the PHSD is based on the dynamical quasiparticle model (DQPM). $\sigma_{\rm el}$ in PHSD decreases with temperature in hadronic phase when approaches towards $T_{c}$ and increases almost linearly for $T_{c} < T$, in the partonic phase after a sudden drop around $T_{c}$. The calculations of quasi-particle (QP) model~\cite{Puglisi:2014pda} are also shown in the figure by solid red line, which match with the PHSD results for QGP phase.

Figure~\ref{eta_s} shows the variation of $\eta/s$ as a function of $\rm T/T_c$. Here, $\rm T_c$ is the critical temperature which is different in different model calculations. The black solid line is the CSPM result and the broken lines are quasiparticle model results~\cite{Thakur:2017hfc}. Here, the dashed line is the result for anisotropic case while the dash-dotted is for isotropic case.  A direct comparison with anisotropic QGP gives a feeling of temperature dependent effect of anisotropy on the discussed observables in figures ~\ref{eta_s} and ~\ref{el}. However, the comparison with the results for isotropic case is only meaningful for CSPM calculations unless the partons are considered as massless. The blue triangle symbols are results of lQCD with (2+1)- dynamical flavours~\cite{Meyer:2007ic,Nakamura:2004sy,Sakai:2007cm}. The black circles are the estimations from dynamical quasiparticle model (DQPM)~\cite{Marty:2013ita}. The red line is the results obtained in QP model~\cite{Puglisi:2014pda}. In CSPM, $\eta/s$ first decreases and after reaching a minimum value, it starts increasing with temperature. Thus, it forms a dip which occurs at $\rm T/T_c$ = 1. The quasi-particle model results~\cite{Thakur:2017hfc} show a similar behaviour but the dip does not occur at critical temperature in this case. We notice that CSPM results are close to the DQPM predictions and stay little higher than the results obtained in the quasiparticle model.  

Recently, the ratio ($\eta$/s)/($\sigma_{\rm el}$/T) has gained a considerable interest in heavy-ion phenomenology~\cite{Puglisi:2014pda}. QGP is expected as a good conductor due to the presence of deconfined color charges. A small value of $\eta/s$ suggests large scattering rates which can damp the conductivity. Since, we know that $\eta/s$ is affected by the gluon-gluon and quark-quark scatterings while $\sigma_{\rm el}$ is only affected by the quark-quark scatterings~\cite{Puglisi:2014pda}. Thus, the ratio between them is important to quantify the contributions from quarks and gluons in various temperature regions. In this work, we have studied this ratio as a function of temperature using CSPM. In figure~\ref{el}, we show the ratio of $\eta$/s and $\sigma_{\rm el}$/T versus $\rm T/T_c$. It is observed that, this ratio behaves in a similar fashion as $\eta/s$. We have also shown the results obtained for the isotropic and anisotropic QGP using a quasi-particle model~\cite{Thakur:2017hfc}. Again, the comparison with the isotropic case is only meaningful. CSPM results are also confronted with the interpolated lattice QCD data~\cite{Puglisi:2014pda} and explain the data within errorbars. The dotted horizontal line is the Ads/CFT calculation~\cite{Puglisi:2014pda} for strongly coupled system. We also show the results obtained in DQPM and QP by the black circles and red line, respectively.

\section{Summary and Outlook}
\label{summary}
In summary, we have developed a method to calculate the electric conductivity of strongly interacting matter using color string percolation approach. We use basically the well-known Drude formula for the estimation of electrical conductivity, which can be obtained after solving the Boltzmann transport equation in relaxation time approximation assuming very small electric fields and no cross effects between heat and electrical conductivity. We see that the CSPM results for the conductivity stays almost constant with increasing temperature in a similar fashion as shown by BAMPS data and matches the results obtained in BAMPS with the fixed strong coupling constant considering elastic cross section only. The CSPM results lie well above the lQCD results for all the temperatures. We have shown $\eta/s$ as a function of $\rm T/T_c$ and compared our results with various quasiparticle models for isotropic and anisotropic cases, lQCD data, DQPM and QP model results. A similar behaviour is found  for CSPM results as shown in lQCD data and other model predictions. But, our results lie above the results obtained from quasiparticle models. CSPM results go inline with that obtained in DQPM. We have also studied the ratio, $(\eta/s)/(\sigma_{\rm el}/T)$ as a function of T, which behaves in a similar manner as $\eta/s$. We have confronted CSPM results with the results obtained in quasiparticle model for isotropic and anisotropic QGP medium, lQCD predictions, estimations from DQPM and QP models. The results obtained for electrical conductivity in CSPM framework validate  the outcomes from  BAMPS calculations with fixed strong coupling constant and fails to explain the predictions of lQCD data.

\section*{ACKNOWLEDGEMENTS}
The authors acknowledge the financial supports from ALICE Project No. SR/MF/PS-01/2014-IITI(G) of Department of Science \& Technology, Government of India.


\begin{thebibliography}{99}

\bibitem{Gyulassy:2004zy} 
  M.~Gyulassy and L.~McLerran,
  Nucl.\ Phys.\ A {\bf 750}, 30 (2005).
  
  
  \bibitem{Gale:2013da} 
  C.~Gale, S.~Jeon and B.~Schenke,
  Int.\ J.\ Mod.\ Phys.\ A {\bf 28}, 1340011 (2013).
  
  
  
  \bibitem{Schenke:2011zz} 
  B.~Schenke, S.~Jeon and C.~Gale,
  J.\ Phys.\ G {\bf 38}, 124169 (2011).
  
  
  \bibitem{Heinz:2013th} 
  U.~Heinz and R.~Snellings,
  Ann.\ Rev.\ Nucl.\ Part.\ Sci.\  {\bf 63}, 123 (2013).
  
  \bibitem{starWhite} J. Adams {\it et al.} (STAR Collaboration), Nucl.\ Phys.\ A {\bf 757}, 102 (2005).
  
  \bibitem{Kubo:1957mj} 
  R.~Kubo,
  J.\ Phys.\ Soc.\ Jap.\  {\bf 12}, 570 (1957).
  
  
  
  \bibitem{Plumari:2012ep} 
  S.~Plumari, A.~Puglisi, F.~Scardina and V.~Greco,
  Phys.\ Rev.\ C {\bf 86}, 054902 (2012).
  
  
  \bibitem{Sasaki:2008fg} 
  C.~Sasaki and K.~Redlich,
  Phys.\ Rev.\ C {\bf 79}, 055207 (2009).
  
  
  \bibitem{Dobado:2008vt} 
  A.~Dobado, F.~J.~Llanes-Estrada and J.~M.~Torres-Rincon,
  Phys.\ Rev.\ D {\bf 79}, 014002 (2009).
  

\bibitem{Chakraborty:2010fr} 
  P.~Chakraborty and J.~I.~Kapusta,
  Phys.\ Rev.\ C {\bf 83}, 014906 (2011).
  
  
  \bibitem{Zhuang:1995uf} 
  P.~Zhuang, J.~Hufner, S.~P.~Klevansky and L.~Neise,
  Phys.\ Rev.\ D {\bf 51}, 3728 (1995).
  
  
  \bibitem{Wiranata:2012br} 
  A.~Wiranata and M.~Prakash,
  Phys.\ Rev.\ C {\bf 85}, 054908 (2012).
  
  
  \bibitem{Ghosh:2014ija} 
  S.~Ghosh,
  Phys.\ Rev.\ C {\bf 90}, 025202 (2014).
  
  
  
  \bibitem{Hirono:2012rt} 
  Y.~Hirono, M.~Hongo and T.~Hirano,
  Phys.\ Rev.\ C {\bf 90}, 021903 (2014).


\bibitem{Tuchin:2013ie} 
  K.~Tuchin,
  Adv.\ High Energy Phys.\  {\bf 2013}, 490495 (2013).
  
  
    \bibitem{Cassing:2013iz} 
  W.~Cassing, O.~Linnyk, T.~Steinert and V.~Ozvenchuk,
  Phys.\ Rev.\ Lett.\  {\bf 110}, 182301 (2013).
  
  
  \bibitem{Fukushima:2008xe} 
  K.~Fukushima, D.~E.~Kharzeev and H.~J.~Warringa,
  Phys.\ Rev.\ D {\bf 78}, 074033 (2008).
  
  
  
 \bibitem{Arnold:2000dr} 
  P.~B.~Arnold, G.~D.~Moore and L.~G.~Yaffe,
  JHEP {\bf 0011}, 001 (2000).
  
  
  \bibitem{Arnold:2003zc} 
  P.~B.~Arnold, G.~D.~Moore and L.~G.~Yaffe,
  JHEP {\bf 0305}, 051 (2003).
  
  
  \bibitem{Gupta:2003zh} 
  S.~Gupta,
  Phys.\ Lett.\ B {\bf 597}, 57 (2004).
  
  
  \bibitem{Aarts:2007wj} 
  G.~Aarts, C.~Allton, J.~Foley, S.~Hands and S.~Kim,
  Phys.\ Rev.\ Lett.\  {\bf 99}, 022002 (2007).
  
  
  \bibitem{Buividovich:2010tn} 
  P.~V.~Buividovich, M.~N.~Chernodub, D.~E.~Kharzeev, T.~Kalaydzhyan, E.~V.~Luschevskaya and M.~I.~Polikarpov,
  Phys.\ Rev.\ Lett.\  {\bf 105}, 132001 (2010).
  
  
  \bibitem{Ding:2010ga} 
  H.-T.~Ding, A.~Francis, O.~Kaczmarek, F.~Karsch, E.~Laermann and W.~Soeldner,
  Phys.\ Rev.\ D {\bf 83}, 034504 (2011).
  
  
  \bibitem{Burnier:2012ts} 
  Y.~Burnier and M.~Laine,
  Eur.\ Phys.\ J.\ C {\bf 72}, 1902 (2012).
  
  \bibitem{Brandt:2012jc} 
  B.~B.~Brandt, A.~Francis, H.~B.~Meyer and H.~Wittig,
  JHEP {\bf 1303}, 100 (2013).
  
  
  \bibitem{Amato:2013naa} 
  A.~Amato, G.~Aarts, C.~Allton, P.~Giudice, S.~Hands and J.~I.~Skullerud,
  Phys.\ Rev.\ Lett.\  {\bf 111}, 172001 (2013).
  
  

  
  
  \bibitem{Steinert:2013fza} 
  T.~Steinert and W.~Cassing,
  Phys.\ Rev.\ C {\bf 89}, 035203 (2014).
  
  
  \bibitem{Puglisi:2014pda} 
  A.~Puglisi, S.~Plumari and V.~Greco,
  Phys.\ Lett.\ B {\bf 751}, 326 (2015).
  
  
  \bibitem{Finazzo:2013efa} 
  S.~I.~Finazzo and J.~Noronha,
  Phys.\ Rev.\ D {\bf 89}, 106008 (2014).
  
  
  \bibitem{Mitra:2016zdw} 
  S.~Mitra and V.~Chandra,
  Phys.\ Rev.\ D {\bf 94}, 034025 (2016).
  
  
  \bibitem{Srivastava:2015via} 
  P.~K.~Srivastava, L.~Thakur and B.~K.~Patra,
  Phys.\ Rev.\ C {\bf 91}, 044903 (2015).
  
  
  \bibitem{Moore:2006qn} 
  G.~D.~Moore and J.~M.~Robert,
  arXiv:hep-ph/0607172.
  
  
  \bibitem{Baym:1997gq} 
  G.~Baym and H.~Heiselberg,
  Phys.\ Rev.\ D {\bf 56}, 5254 (1997).
  
  
  \bibitem{FernandezFraile:2005ka} 
  D.~Fernandez-Fraile and A.~Gomez Nicola,
  Phys.\ Rev.\ D {\bf 73}, 045025 (2006).
  
  \bibitem{Armesto:1996kt} 
  N.~Armesto, M.~A.~Braun, E.~G.~Ferreiro and C.~Pajares,
  Phys.\ Rev.\ Lett.\  {\bf 77}, 3736 (1996).
  
  
  \bibitem{Nardi:1998qb} 
  M.~Nardi and H.~Satz,
  Phys.\ Lett.\ B {\bf 442}, 14 (1998).
  
  \bibitem{Braun:1999hv} 
  M.~A.~Braun and C.~Pajares,
  Eur.\ Phys.\ J.\ C {\bf 16}, 349 (2000).
  
  
  \bibitem{Braun:1997ch} 
  M.~A.~Braun, C.~Pajares and J.~Ranft,
  Int.\ J.\ Mod.\ Phys.\ A {\bf 14}, 2689 (1999).
  
  
  \bibitem{Braun:2000hd} 
  M.~A.~Braun and C.~Pajares,
  Phys.\ Rev.\ Lett.\  {\bf 85}, 4864 (2000).
  
  
  
\bibitem{Phyreport}
  M.A.~Braun {\it et al.} 
 Phys.\ Reports.\ {\bf 509}, 1 (2015).


 \bibitem{PLB642}
  J.~Dias de Deus and C.~Pajares,
  Phys.\ Lett.\ B {\bf 642}, 455 (2006).
  
  \bibitem{Gutay:2015cba} 
  L.~J.~Gutay {\it et al.},
  Int.\ J.\ Mod.\ Phys.\ E {\bf 24}, 1550101 (2015).
  
  \bibitem{Hirsch:2018pqm} 
  A.~S.~Hirsch, C.~Pajares, R.~P.~Scharenberg and B.~K.~Srivastava,
  arXiv:1803.02301 [hep-ph].


\bibitem{Sahoo:2018dcz} 
  P.~Sahoo, S.~De, S.~K.~Tiwari and R.~Sahoo,
  Eur.\ Phys.\ J.\ A {\bf 54}, 136 (2018)


\bibitem{Sahoo:2017umy} 
  P.~Sahoo, S.~K.~Tiwari, S.~De, R.~Sahoo, R.~P.~Scharenberg and B.~K.~Srivastava,
  arXiv:1708.06689 [hep-ph].



\bibitem{Braun:2015eoa} 
  M.~A.~Braun, J.~Dias de Deus, A.~S.~Hirsch, C.~Pajares, R.~P.~Scharenberg and B.~K.~Srivastava,
  Phys.\ Rept.\  {\bf 599}, 1 (2015).
    
  
  
  \bibitem{Anderson}
  J. Anderson and H. Witting, Physica (Utrecht) {\bf 74}, 466 (1974).
  
  
%
%
%
  
  \bibitem{Greif:2014oia} 
  M.~Greif, I.~Bouras, C.~Greiner and Z.~Xu,
  Phys.\ Rev.\ D {\bf 90}, 094014 (2014).
  
  
  \bibitem{Greif:2016skc} 
  M.~Greif, C.~Greiner and G.~S.~Denicol,
  Phys.\ Rev.\ D {\bf 93}, 096012 (2016).
  Erratum: [Phys.\ Rev.\ D {\bf 96}, 059902 (2017).]

  
  
  \bibitem{CaronHuot:2006te} 
  S.~Caron-Huot, P.~Kovtun, G.~D.~Moore, A.~Starinets and L.~G.~Yaffe,
  JHEP {\bf 0612}, 015 (2006).
  
  
  \bibitem{Puglisi:2014sha} 
  A.~Puglisi, S.~Plumari and V.~Greco,
  Phys.\ Rev.\ D {\bf 90}, 114009 (2014).

  
  \bibitem{DiasdeDeus:2012uc} 
  J.~Dias de Deus {\it et al.},
  Eur.\ Phys.\ J.\ C {\bf 72}, 2123 (2012).
  
  
  
  \bibitem{Danielewicz:1984ww} 
  P.~Danielewicz and M.~Gyulassy,
  Phys.\ Rev.\ D {\bf 31}, 53 (1985).
  
  
  \bibitem{Hirano:2005wx} 
  T.~Hirano and M.~Gyulassy,
  Nucl.\ Phys.\ A {\bf 769}, 71 (2006).





  
\bibitem{Aarts:2014nba} 
  G.~Aarts, C.~Allton, A.~Amato, P.~Giudice, S.~Hands and J.~I.~Skullerud,
  JHEP {\bf 1502}, 186 (2015).
  
  
  
  
  
  
  
  \bibitem{Thakur:2017hfc} 
  L.~Thakur, P.~K.~Srivastava, G.~P.~Kadam, M.~George and H.~Mishra,
  Phys.\ Rev.\ D {\bf 95}, 096009 (2017).
  
  
  \bibitem{Meyer:2007ic} 
  H.~B.~Meyer,
  Phys.\ Rev.\ D {\bf 76}, 101701 (2007).
  
  
  \bibitem{Nakamura:2004sy} 
  A.~Nakamura and S.~Sakai,
  Phys.\ Rev.\ Lett.\  {\bf 94}, 072305 (2005).
  
  
  \bibitem{Sakai:2007cm} 
  S.~Sakai and A.~Nakamura,
  PoS LATTICE {\bf 2007}, 221 (2007).
  
  
  \bibitem{Marty:2013ita} 
  R.~Marty, E.~Bratkovskaya, W.~Cassing, J.~Aichelin and H.~Berrehrah,
  Phys.\ Rev.\ C {\bf 88}, 045204 (2013).
  
  



  
  
   
\end{thebibliography}
\end{document}